\documentclass[prd,nofootinbib,preprint,superscriptaddress]{revtex4}
\pdfoutput=1
\usepackage[T1]{fontenc}
\usepackage{amsmath,amssymb}
\usepackage{epsfig}
\usepackage{subfigure}
\usepackage{float}
\usepackage{graphicx}
\usepackage[usenames,dvipsnames]{color}
\usepackage{bigints}
\usepackage{slashed}
\usepackage{multirow}
\usepackage[colorlinks,citecolor=blue]{hyperref}
\usepackage{pdfpages}
\usepackage{color}
\usepackage{comment}

 \catcode`\@=11
 \def\lsim{\mathrel{\mathpalette\@versim<}}
 \def\gsim{\mathrel{\mathpalette\@versim>}}
 \def\@versim#1#2{\vcenter{\offinterlineskip
 \ialign{$\m@th#1\hfil##\hfil$\crcr#2\crcr\sim\crcr } }}
 \catcode`\@=12

 \catcode`@=12
\begin{document}
 \thispagestyle{empty}
 \begin{flushright}
 UCRHEP-T621\\
 Sep 2022\
 \end{flushright}
 

 \vspace{0.6in}
 \begin{center}
 {\LARGE \bf Dark $SU(2)$ Gauge Symmetry\\
and Scotogenic Dirac Neutrinos\\}
 \vspace{1.5in}
{\bf Debasish Borah \\}
 \vspace{0.1in}
{\sl Department of Physics, Indian Institute of Technology
Guwahati, Assam 781039, India \\}
 
\vspace{0.2in}
 {\bf Ernest Ma\\}
 \vspace{0.1in}
{\sl Department of Physics and Astronomy,\\ 
University of California, Riverside, California 92521, USA\\}

 \vspace{0.2in}
{\bf Dibyendu Nanda \\}
 \vspace{0.1in}
{\sl School of Physics, Korea Institute for Advanced Study, Seoul 02455, Korea \\}

\end{center}
 \vspace{1.0in}

\centerline{\bf Abstract}
\vspace{0.1in}
Dark matter is postulated to transform under the non-Abelian 
$SU(2)_D$ gauge symmetry.  Its connection to Dirac neutrino masses is explored.


\newpage
\baselineskip 24pt

\noindent \underline{\it Introduction}~:~ 
The existence of dark matter~\cite{Young:2016ala} is well established, but 
its nature is still unknown.  It is however very likely to be stabilized by an 
unbroken symmetry, such as $Z_2$ dark parity, under which it is odd 
and all particles of the Standard Model (SM) are even.  If it has 
interactions with Majorana neutrinos, it is also derivable~\cite{Ma:2015xla} 
from lepton parity, i.e. $\pi_D = \pi_L (-1)^{2j}$, where $\pi_L = (-1)^L$ 
and $j$ is the intrinsic spin of the particle.  If the dark symmetry is 
global $U(1)_D$, which may be the remnant~\cite{Ma:2013yga} of an $U(1)$ gauge 
symmetry, it is possibly connected to Dirac 
neutrinos~\cite{Ma:2019coj,Ma:2022uhi}.

To explore further the origin of dark symmetry, the non-Abelian gauge 
symmetry $SU(2)_D$ may be considered~\cite{Hambye:2008bq,Ma:2018ouq}.  
In this paper, a model is studied where $SU(2)_D$ is spontaneously broken 
to retain a global $SU(2)$ symmetry, with additional fermions and scalars 
which link them to the leptons of the Standard Model (SM) of particle 
interactions. Two Dirac neutrino masses are generated in one loop from 
these $SU(2)_D$ dark fermions and scalars.  

\noindent \underline{\it Model}~:~ 
The particle content of this model is shown in Table \ref{tab1} and \ref{tab2}.  The scalar 
doublet $(\zeta_1,\zeta_2)$ breaks the gauge $SU(2)_D$ symmetry to 
global $SU(2)_D$.  Two dark fermion doublets $(\psi_{L1},\psi_{L2})$,
$(\psi_{R1},\psi_{R2})$ are added, together with one scalar bidoublet 
$(\tilde{\nu}_{L1},\tilde{l}_{L2};\tilde{\nu}_{L2},\tilde{l}_{L2})$ 
and one scalar doublet $(\tilde{\nu}_{R1},\tilde{\nu}_{R2})$.
\begin{table}[tbh]
\centering
\begin{tabular}{|c|c|c|c|c|c|c|}
\hline
fermion & $SU(3)_C$ & $SU(2)_L$ & $U(1)_Y$ & $SU(2)_D$ & $L$ & $Z_2$ \\
\hline
$(\nu,l)_L$ & 1 & $2$ & $-1/2$ & $1$ & $1$ & $+$ \\ 
$l_R$ & 1 & $1$ & $-1$ & 1 & 1 & $+$ \\ 
\hline
$\nu_R$ & 1 & $1$ & $0$ & 1 & 1 & $-$ \\ 
\hline
$(\psi_{L1}, \psi_{L2})$ & 1 & 1 & 0 & 2 & 0 & + \\
$(\psi_{R1}, \psi_{R2})$ & 1 & 1 & 0 & 2 & 0 & + \\
\hline
\end{tabular}
\caption{Relevant fermion content the dark $SU(2)_D$ model.}
\label{tab1}
\end{table}

\begin{table}[tbh]
\centering
\begin{tabular}{|c|c|c|c|c|c|c|}
\hline
scalar & $SU(3)_C$ & $SU(2)_L$ & $U(1)_Y$ & $SU(2)_D$ & $L$ & $Z_2$ \\
\hline
$(\phi^+,\phi^0)$ & 1 & $2$ & $1/2$ & $1$ & $0$ & $+$ \\ \hline
$(\zeta_1, \zeta_2)$ & 1 & 1 & 0 & 2 & 0 & + \\
\hline
$(\tilde{\nu}_{L1},\tilde{l}_{L1};\tilde{\nu}_{L2},\tilde{l}_{L2})$ 
& 1 & $2$ & $-1/2$ & $2$ & $1$ & $+$ \\ 
\hline
$(\tilde{\nu}_{R1}, \tilde{\nu}_{R2})$ & 1 & $1$ & $0$ & 2 & 1 & $-$ \\ 
\hline
\end{tabular}
\caption{Scalar content of the dark $SU(2)_D$ model.}
\label{tab2}
\end{table}
Note that one chiral $SU(2)_D$ fermion doublet is by itself anomaly-free, 
but is massless (because the $SU(2)_D$ invariant bilinear term 
$\psi_1 \psi_2 - \psi_2 \psi_1$ term is antisymmetric and identically 
zero) unless there is also a scalar $SU(2)_D$ triplet, as in 
Ref.~\cite{Ma:2018ouq}.  In the case of two chiral $SU(2)_D$ fermion 
doublets as in this model, the invariant 
$\bar{\psi}_{L1} \psi_{R1} + \bar{\psi}_{L2} \psi_{R2}$ is a nonzero 
mass term.  If there are three chiral $SU(2)_D$ fermion doublets, 
it is easily shown that two will pair up as in this model, and the other 
will remain massless.

As shown in Table \ref{tab1} and \ref{tab2}, lepton number $L$ is imposed and required to be 
conserved by all terms in the Lagrangian.  The $Z_2$ symmetry applies 
to all dimension-four terms of the Lagrangian, but is broken by soft 
dimension-three terms.  It serves the purpose of forbidding the coupling of 
the right-handed neutrino $\nu_R$ to $\nu_L$ through $\phi^0$.  Hence 
neutrinos are massless at tree level.  They will pick up Dirac masses in 
one loop through the soft breaking of $Z_2$ to be discussed.  Whereas 
there are three families of $\nu_L$, there are only two of $\nu_R$. 
If more than two are assumed, only two linear combinations will pick 
up masses because only two $SU(2)_D$ fermion doublets are postulated. 
Therefore, one neutrino must be massless in this model, which is of 
course allowed by neutrino oscillations where only mass-squared differences 
are measured.

The spontaneous breaking of the $SU(2)_D$ gauge symmetry comes from the 
$SU(2)_D$ scalar doublet $\zeta$.  Without loss of generality, 
$\langle \zeta_2 \rangle = v_D$ may be chosen.  As a result, all three 
$SU(2)_D$ gauge bosons $W_D$ acquire the same mass 
$m_{W_D} = g_D v_D/\sqrt{2}$, and the one physical dark Higgs boson is 
$h_D = \sqrt{2} Re(\zeta_2)$.
The Higgs potential consisting of $\Phi$ and $\zeta$ is  
\begin{equation}
V = m_1^2 \Phi^\dagger \Phi + m_2^2 \zeta^\dagger \zeta + {1 \over 2} 
\lambda_1 (\Phi^\dagger \Phi)^2 + {1 \over 2} \lambda_2 
(\zeta^\dagger \zeta)^2 + \lambda_3 (\Phi^\dagger \Phi)(\zeta^\dagger \zeta).
\end{equation}
The other scalars, i.e. the bidoublet 
$(\tilde{\nu}_{L1}, \tilde{l}_{L1};\tilde{\nu}_{L2}, \tilde{l}_{L2})$ 
and the doublet  $(\tilde{\nu}_{R1}, \tilde{\nu}_{R2})$ have $L=1$ 
and will not develop vacuum expectation values.  Their quartic interactions 
with $\Phi$ and $\zeta$ are products of bilinear invariants, which are 
easily written out and are not very illuminating or relevant.  There are 
however trilinear terms which break $Z_2$ softly and are important for 
generating one-loop Dirac neutrino masses to be discussed. 
Let $\langle \phi^0 \rangle = v$ and $\langle \zeta \rangle = v_D$, then
\begin{equation}
v^2 = {-\lambda_2 m_1^2 + \lambda_3 m_2^2 \over \lambda_1 \lambda_2 
- \lambda_3^2}, ~~~ v_D^2 = {-\lambda_1 m_2^2 + \lambda_3 m_1^2 \over 
\lambda_1 \lambda_2 - \lambda_3^2}.
\end{equation}
The only physical scalars are $h_D$ and the SM $h=\sqrt{2}Re(\phi^0)$, 
resulting in 
\begin{eqnarray}
V &=& \lambda_1 v^2 h^2 + \lambda_2 v_D^2 h_D^2 + 2 \lambda_3 v v_D h h_D + 
{1 \over \sqrt{2}} \lambda_1 v h^3 + {1 \over \sqrt{2}} \lambda_2 v_D h_D^3 
\nonumber \\ &+& {1 \over \sqrt{2}} \lambda_3 v h h_D^2 + {1 \over \sqrt{2}} 
\lambda_3 v_D h_D h^2 + {1 \over 8} \lambda_1 h^4 + {1 \over 8} \lambda_2 
h_D^4 + {1 \over 4} \lambda_3 h^2 h_D^2.
\end{eqnarray}
Hence $h_D$ mixes with $h$ according to
\begin{equation}
M^2_{h h_D} = \begin{pmatrix} 2 \lambda_1 v^2 & 2 \lambda_3 v v_D \cr 2 \lambda_3 v 
v_D & 2 \lambda_2 v_D^2 \end{pmatrix},
\end{equation}
and may decay to SM fermions through the $h$ Yukawa couplings.  Note that 
$h_D$ does not couple directly to $\bar{\psi}\psi$.

\begin{figure}[htb]
 \vspace*{-5cm}
 \hspace*{-5cm}
 \includegraphics[scale=1.0]{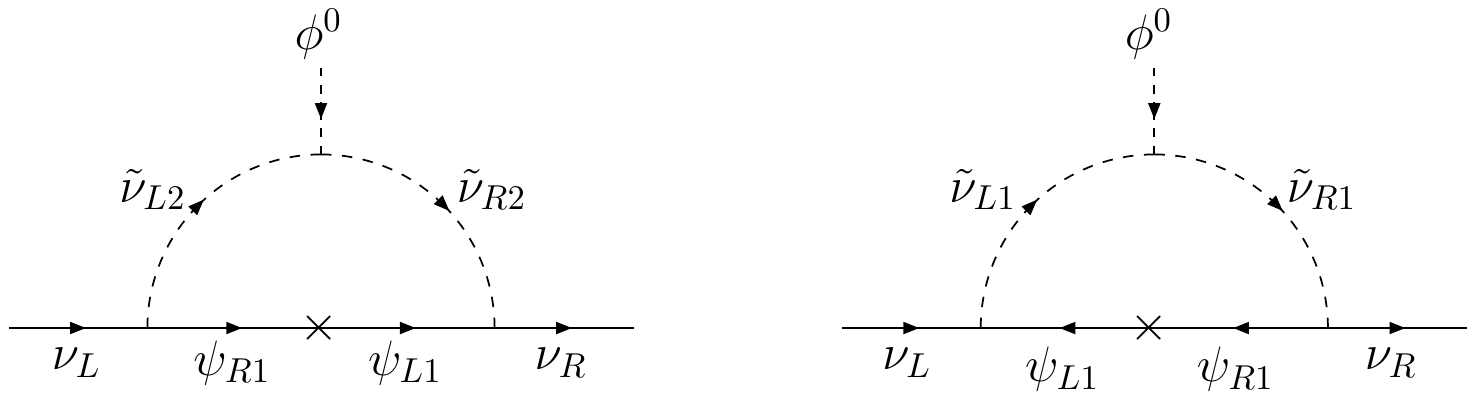}
 \vspace*{-20.5cm}
 \caption{Scotogenic Dirac neutrino mass.}
 \label{fig:numass}
 \end{figure}

\noindent \underline{\it Scotogenic Dirac Neutrino Mass}~:~
As pointed out earlier, the $Z_2$ symmetry of Table \ref{tab1}, \ref{tab2} is required for all 
dimension-four couplings.  Hence $\nu_R$ cannot couple to $\nu_L$ through 
$\phi^0$, but they do couple to the dark fermions with $L=0$ and dark 
scalars with $L=1$, i.e.
\begin{eqnarray}
{\cal L}_Y &=& f^L_{i1}[\nu_{Li} (\psi_{L1} \tilde{\nu}^*_{L1} + 
\psi_{L2} \tilde{\nu}^*_{L2}) + l_{Li} (\psi_{L1} \tilde{l}^*_{L1} + 
\psi_{L2} \tilde{l}^*_{L2})] \nonumber \\ 
&+& f^L_{i2}[\bar{\nu}_{Li} (\psi_{R1} \tilde{\nu}_{L2} - 
\psi_{R2} \tilde{\nu}_{L1}) + \bar{l}_{Li} (\psi_{R1} \tilde{l}_{L2} - 
\psi_{R2} \tilde{l}_{L1})] \nonumber \\ 
&+& f^R_{i1} \nu_{Ri} (\psi_{R1} \tilde{\nu}^*_{R1} + \psi_{R2} 
\tilde{\nu}^*_{R2}) + f^R_{i2} \bar{\nu}_{Ri} (\psi_{L1} \tilde{\nu}_{R2} 
- \psi_{L2} \tilde{\nu}_{R1}) + H.c.
\end{eqnarray} 
Thus there are two connections linking $\nu_L$ to $\nu_R$, as shown in 
Fig. \ref{fig:numass}.  It is understood that they are accompanied by two more diagrams 
which render the vertices invariant under $SU(2)_D$ as given in ${\cal L}_Y$. 
The $Z_2$ symmetry is broken softly by the trilinear term $(\tilde{\nu}_{R1}^* \tilde{\nu}_{L1}+\tilde{\nu}_{R2}^* \tilde{\nu}_{L2})\phi^0$. 
Note also that the lepton number $L$ assignments of Table \ref{tab1}, \ref{tab2} are mandatory 
because of the $SU(2)_D$ gauge symmetry which forces the $\psi$ fermions 
to have $L=0$.
 
The scalar $\tilde{\nu}_L$ and $\tilde{\nu}_R$ doublets mix through their 
coupling to $\phi^0$.  Let their mass eigenstates be 
\begin{equation}
x = \tilde{\nu}_L \cos \theta - \tilde{\nu}_R \sin \theta, ~~~ 
y = \tilde{\nu}_L \sin \theta + \tilde{\nu}_R \cos \theta,
\end{equation} 
then the radiative Dirac neutrino mass matrix is given by
\begin{equation}
({\cal M}_\nu)_{ij} = {\sin \theta \cos \theta ~m_\psi \over 8 \pi}  
[F(m_x^2,M_\psi^2)-F(m_y^2,M_\psi^2)] \sum_{k=1,2} f^L_{ik} f^R_{kj},
\end{equation}
where $F(a,b)=a \ln(a/b)/(a-b)$.  Thus two neutrinos are massive and one 
is massless in this model, allowing two mass-squared differences as 
observed in neutrino oscillations.

\begin{figure}[htb]
 \vspace*{-5cm}
 \hspace*{-3cm}
 \includegraphics[scale=1.0]{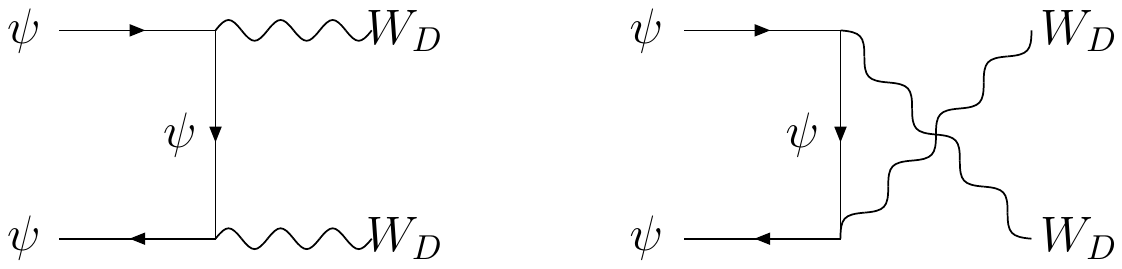}
 \vspace*{-21.5cm}
 \caption{Annihilation of $\psi \bar{\psi}$ to $W_D W_D$.}
 \label{fig:psiann}
 \end{figure}

\begin{figure}[htb]
 \vspace*{-5cm}
 \hspace*{-3cm}
 \includegraphics[scale=1.0]{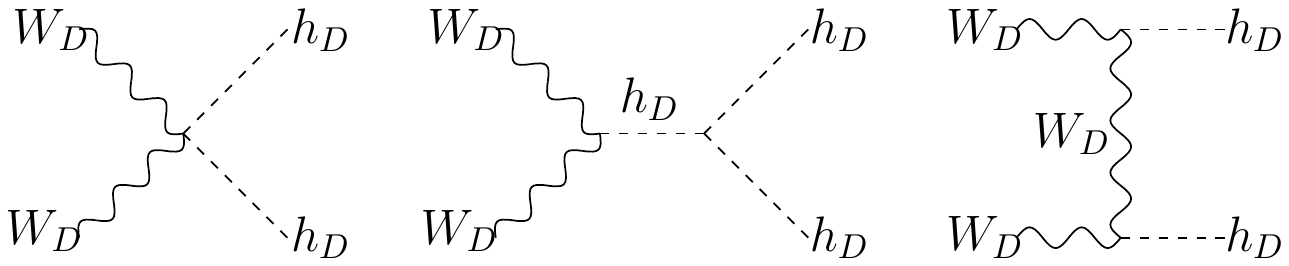}
 \vspace*{-21.5cm}
 \caption{Annihilation of $W_D W_D$ to $h_D h_D$.}
 \label{fig:wdann}
 \end{figure}

\noindent \underline{\it Freeze-Out Scenario}~:~
The dark sector consists of the three $W_D$ gauge bosons, the fermion 
$SU(2)_D$ doublet $(\psi_1,\psi_2)$, 
and the scalar $SU(2)_D$ doublets $(\tilde{l}_{L1},\tilde{l}_{L2})$, as 
well as the linear combinations of $(\tilde{\nu}_{L1},\tilde{\nu}_{L2})$ 
and $(\tilde{\nu}_{R1},\tilde{\nu}_{R2})$ with masses $m_{x,y}$. 
Assuming that $m_{x,y} > M_\psi > M_{W_D}$, then both $\psi$ and $W_D$ are 
stable because the former is a doublet and the latter a triplet under 
unbroken global $SU(2)_D$.  Their relic densities from thermal freeze-out 
are determined by $\psi \bar{\psi}$ annihilating to $W_D W_D$ as shown in 
Fig. \ref{fig:psiann}, 
and the subsequent annihilation of $W_D W_D$ to $h_D h_D$ as 
shown in Fig. \ref{fig:wdann}. 

The respective cross sections $\times$ relative velocity are~\cite{Ma:2021szi}
\begin{equation}
\sigma(\psi \bar{\psi} \to W_D W_D) v_{\rm rel} = {3g_D^4 \over 256 \pi M_\psi^2} 
\left(1-{M^2_{W_D} \over M^2_\psi}\right)^{3 \over 2} \left(1-{M^2_{W_D} \over 
2M^2_\psi}\right)^{-2},
\end{equation}
and~\cite{Ma:2017ucp}
\begin{equation}
\sigma(W_D W_D \to h_D h_D) v_{\rm rel} = {g_D^4 \sqrt{1-r} \over 576 \pi 
M^2_{W_D}}[3A^2+2AB(1-r)+B^2(1-r)^2],
\end{equation}
where $r=M^2_{h_D}/M^2_{W_D}$ and
\begin{equation}
A={1 \over 2} \left[1+{3r \over 4-r}-{4 \over 2-r} \right], ~~~ 
B={2 \over 2-r}.
\end{equation}

\begin{figure}[htb!]
    \centering
\includegraphics[scale=0.47]{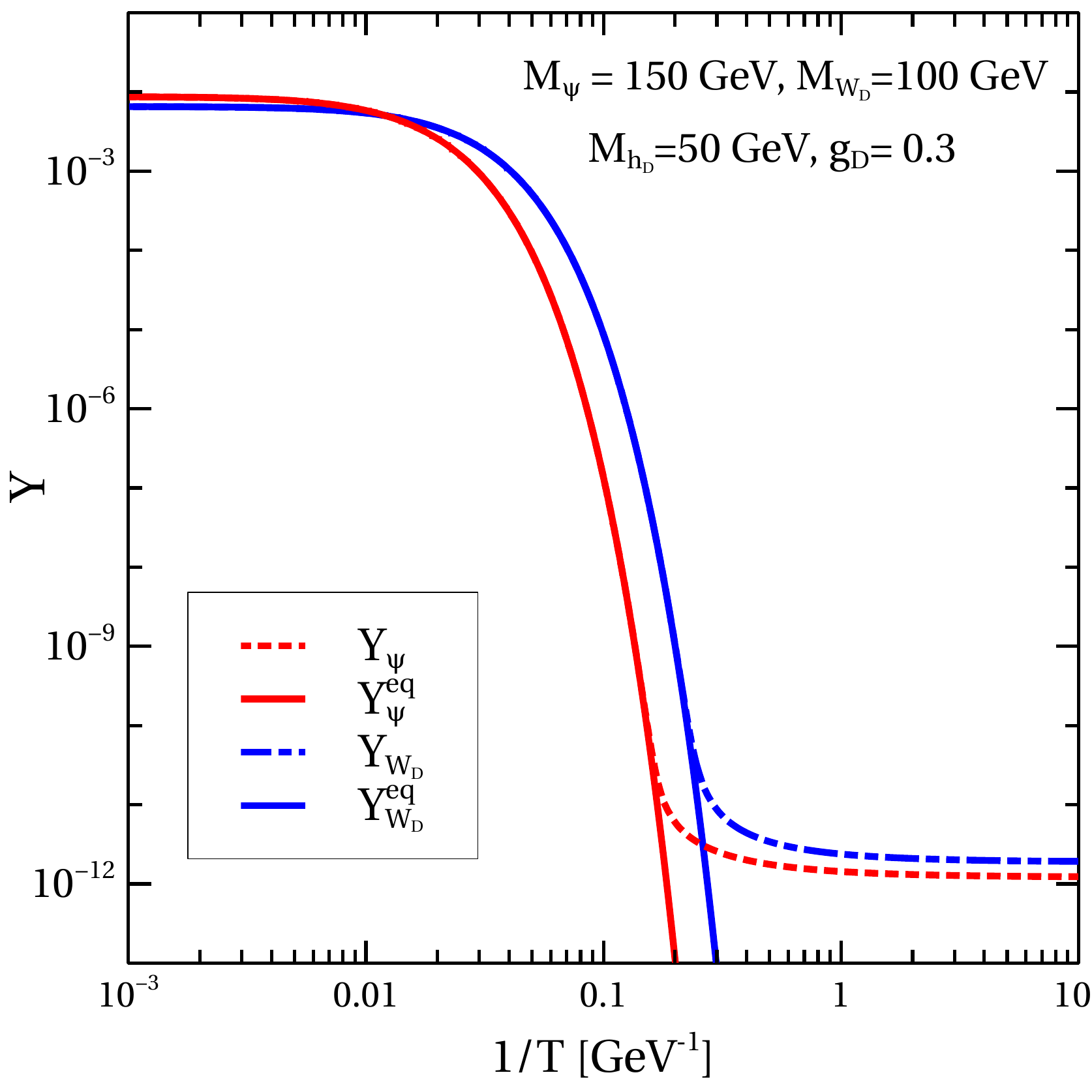}\,
\includegraphics[scale=0.47]{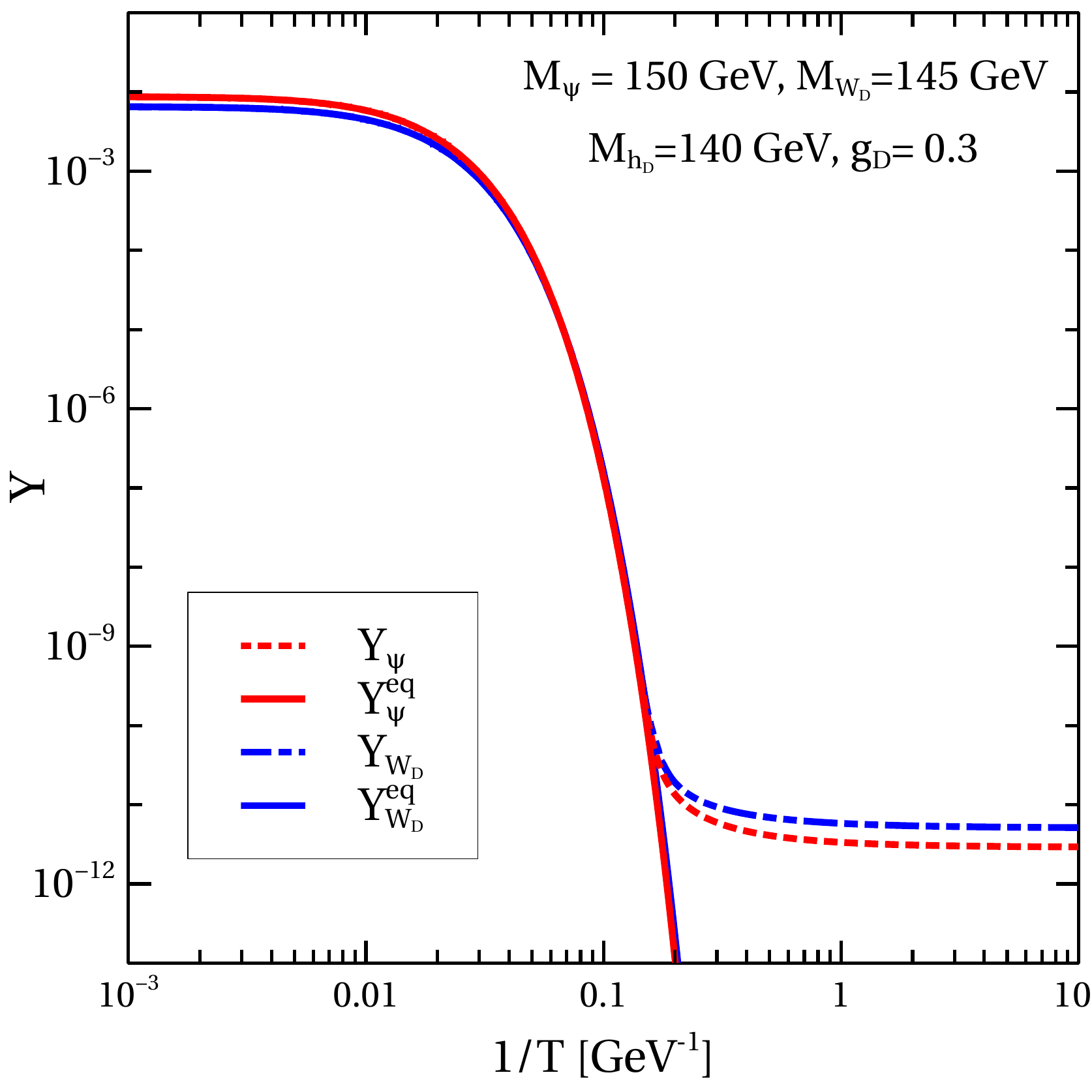}
    \caption{Evolution of comoving number densities of both $\psi$, $W_D$ as a function of temperature  for different benchmark values of the other parameters. }
    \label{fig:com:psi}
\end{figure}
\begin{figure}[htb!]
\includegraphics[scale=0.47]{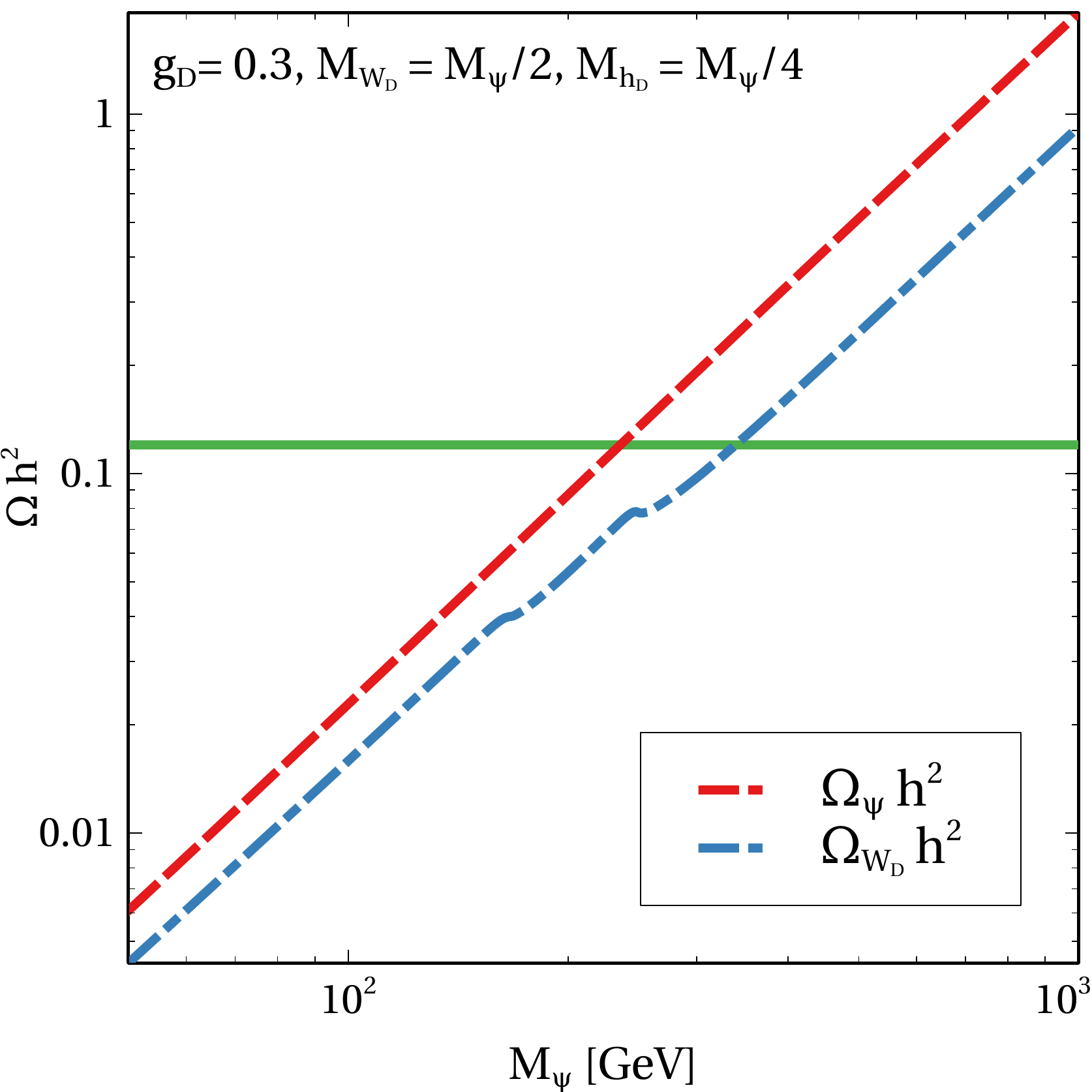}\,
\includegraphics[scale=0.47]{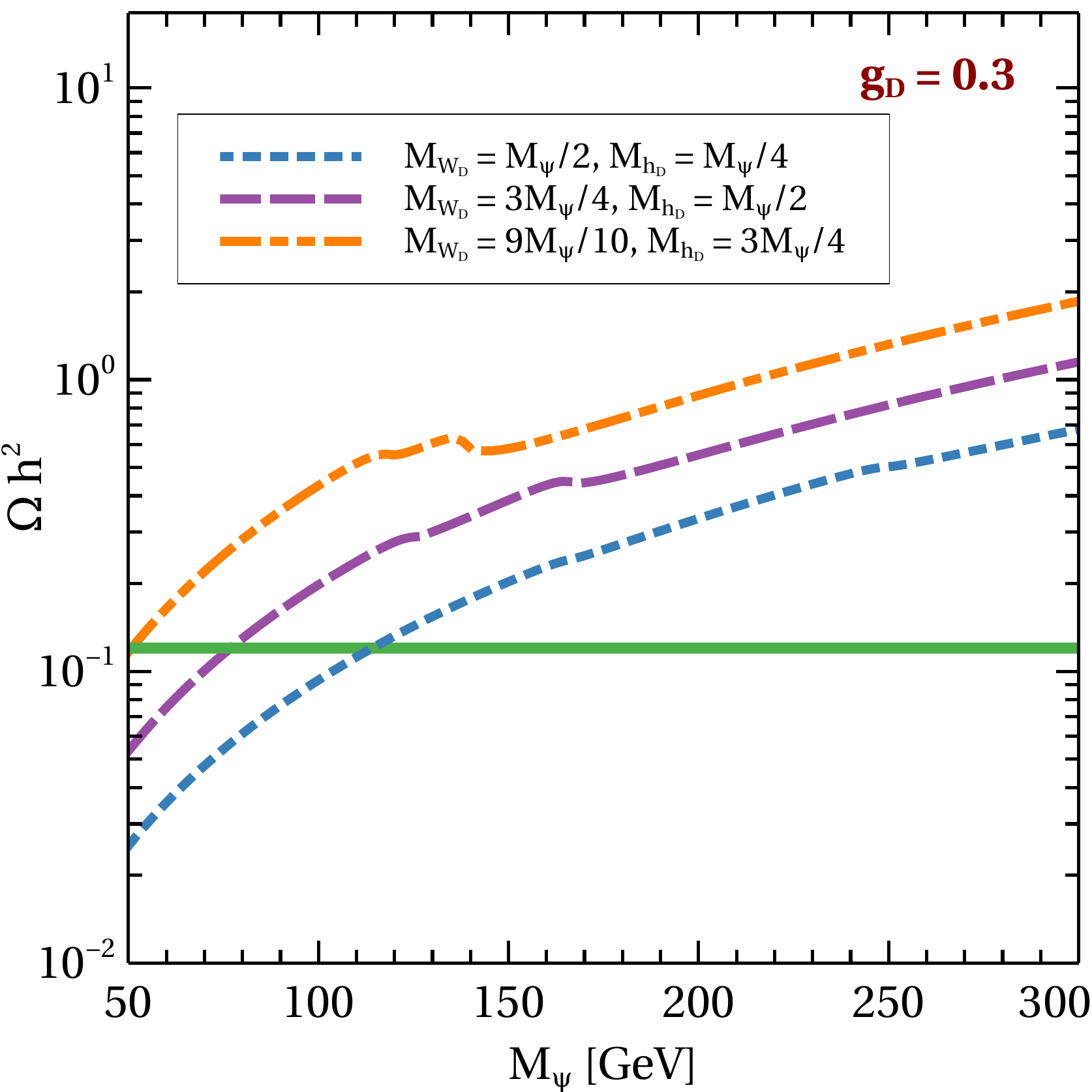}
\caption{Left panel: Relic density of the individual dark matter component as a function of $M_{\psi}$. Right panel: Total relic density of dark matter as a function of $M_{\psi}$ for different benchmark values of other parameters. } 
\label{fig:relic:line}
\end{figure}

In order to discuss the relic density of two-component dark matter \cite{Cao:2007fy, Zurek:2008qg, Liu:2011aa, Belanger:2011ww, Adulpravitchai:2011ei} quantitatively, we first write the coupled Boltzmann equations (BEs) for comoving number densities of $W_D$ and $\psi$ respectively. Choosing the variable to be $x_i=(M_i/T)$ with $i= 1,2$ for $W_D, \psi$ respectively, the BEs in the limit of two stable dark matter candidates with the dominant annihilations specified above, can be written as
\begin{eqnarray}\nonumber
\frac{dY_{W_D}}{dx_1} &=& \frac{\beta s}{H x_1}\bigg(-\left<\sigma (W_D W_D \rightarrow h_D h_D)v_{\rm rel} \right>\left[Y_{W_D}^2- \left(Y_{W_D}^{\rm eq}\right)^2 \right]\\  &&   + \left<\sigma (\psi \psi \rightarrow W_D W_D)v_{\rm rel} \right>\left[Y_{\psi}^2 - \frac{\left(Y_{\psi}^{\rm eq}\right)^2}{\left(Y_{W_D}^{\rm eq}\right)^2} Y_{W_D}^2 \right]\bigg)\label{BE:WD:Y}\\
\frac{dY_{\psi}}{dx_2} &=& \frac{\beta s}{H x_2} \left(- \left<\sigma (\psi \psi \rightarrow W_D W_D)v_{\rm rel} \right>\left[Y_{\psi}^2 - \frac{\left(Y_{\psi}^{\rm eq}\right)^2}{\left(Y_{W_D}^{\rm eq}\right)^2} Y_{W_D}^2 \right]\right),
\label{BE:psi:Y}
\end{eqnarray}
where $$\beta (T)= 1 + \frac{1}{3} \frac{T}{g_s(T)}\frac{d g_s(T)}{dT}$$ and $H$ is the Hubble parameter. The respective cross-sections $(\sigma v_{\rm rel})$ are given above while $ \langle \sigma v_{\rm rel} \rangle $ is the thermally averaged annihilation cross-section \cite{Gondolo:1990dk}.
The first term of the right hand side of Eq.~\eqref{BE:WD:Y} represents the effect of the annihilation of two $W_D$ to two dark Higgs ($h_D$) whereas the second term shows the conversion process between $\psi$ and $W_D$. On the other hand, the right hand side of Eq.~\eqref{BE:psi:Y}, contains the conversion term only as $\psi$ dominantly annihilates into the other dark matter, namely $W_D$. In order to calculate the thermally averaged annihilation cross-sections and solve the above BEs numerically, we use \texttt{micrOMEGAs} \cite{Belanger:2014vza}. In Fig.~\ref{fig:com:psi}, we have shown the evolution of the comoving number densities of the two DM candidates as a function of temperature while fixing relevant parameters to some benchmark values. For both the plot we have fixed the mass of $\psi$ and changed the masses of the other dark sector particles. Since $M_\psi > M_{W_D}$ in the limit of two stable DM candidates, we always have conversion of heavier DM $\psi$ into lighter DM $W_D$. Also, since $\psi$ annihilates only into $W_D$, the final comoving number density of $\psi$ decreases when the mass splitting between $\psi$ and $W_D$ is small and vice versa. This is because, for larger mass difference, the cross-section for $\psi \psi \rightarrow W_D W_D$ gets a phase-space enhancement leading to decrease in the relic density of $\psi$. In Fig.~\ref{fig:relic:line}, we have shown the dependence of relic density as a function of $M_{\psi}$ for fixed value of $SU(2)_D$ gauge coupling $g_D = 0.3$ and specific mass relations between dark matter candidates and the dark Higgs $h_D$. The left panel shows the contribution coming from individual dark matter components for $M_{W_D}=M_{\psi}/2$ and $M_{h_D}=M_{\psi}/4$. One can clearly notice that the $\psi$ contributes more to the relic density than the $W_D$. This is because, being the heaviest dark sector particle with only one annihilation channel affecting its relic density, $\psi$ decouples earlier than $W_D$ with the latter being in thermal equilibrium for longer duration by virtue of its multiple annihilation channels. The efficient annihilation of $W_D$ till late epochs leads to a lower relic density.
The right panel of Fig.~\ref{fig:relic:line} shows total relic density as function of $M_\psi$ where the other parameters have been fixed as shown in the figure itself. As the mass difference between different dark sector particles is reduced, the total relic increases due to inefficient annihilation rates as expected.

Apart from the relic density constraints, this model can also be tested in the ongoing and future direct detection experiments \cite{LUX-ZEPLIN:2022qhg}. As for direct detection, $\psi$ has no tree-level interactions with quarks, whereas $W_D$ could do so through $h-h_D$ mixing.  The typical 
constraint~\cite{Ma:2021roh} on $\lambda_3$ is that it should be less than about $10^{-4}$. On the other hand, $W_D$ can have tree level spin-independent scattering off nucleons via $h_D$-SM Higgs mixing. Due to the freedom in choosing this scalar mixing, as long as it is large enough to thermalise the dark sector, one can keep such tree level DM-nucleon cross section under control. The corresponding cross-section can be written as 
\begin{equation}
    \sigma_{\rm SI}^{W_D} = \frac{g_D^2 f_p^2 m_n^4}{2\pi (M_{W_D} + m_n)^2} \frac{ \xi^2 M_{W_D}^2 (m_h^2 - M_{h_D}^2)^2}{m_h^4 M_{h_D}^4 v^2},
\end{equation}
where $m_n$ is the mass of the nucleon, $f_p$ and $\xi$ are form factor and dark Higgs-SM Higgs mixing parameter respectively. Such small $h_D$-SM Higgs mixing also keep the invisible decay rate of SM Higgs within control, for light $h_D$ below SM-Higgs mass threshold. The latest constraints on such invisible Higgs decay is ${\rm BR}_{\rm h \rightarrow {\rm inv}} < 14.6\%$ \cite{ATLAS:2022yvh} and ${\rm BR}_{\rm h \rightarrow {\rm inv}} < 18\%$ \cite{CMS:2022qva}. For $M_{h_D} < m_h/2$, this rules out $h_D$-SM Higgs portal coupling $\lambda_3$ upto $\mathcal{O}(10^{-2})$. Other relevant bounds on such mixing can arise from electroweak precision data \cite{Robens:2015gla,Chalons:2016jeu, Lopez-Val:2014jva} as well as direct collider searches  \cite{Khachatryan:2015cwa,Strassler:2006ri}. The chosen value of mixing parameter $\xi=0.001$ automatically satisfies these bounds for $h_D$ mass above a few tens of GeV.

\begin{figure}[htb!]
\includegraphics[scale=0.47]{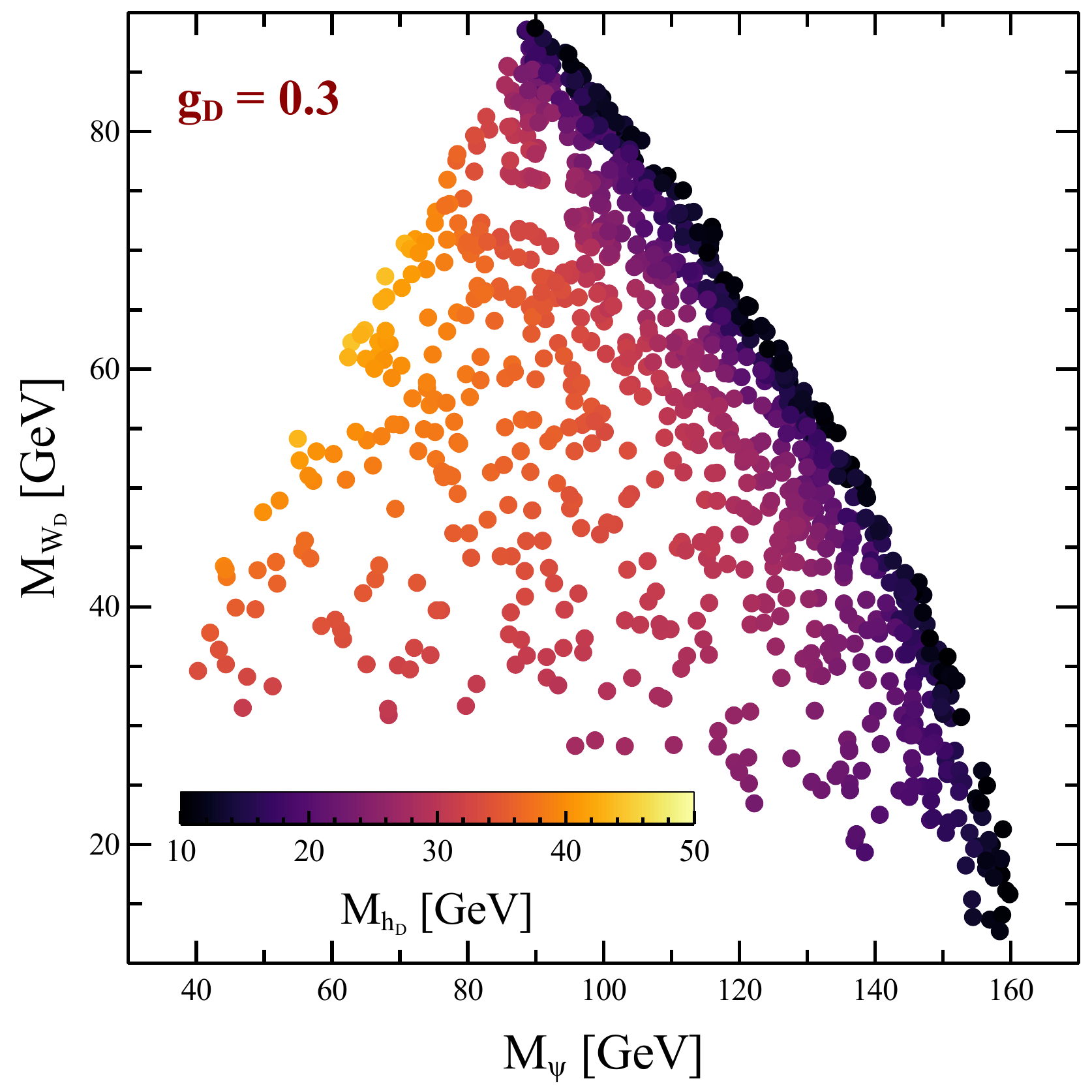}\,
\includegraphics[scale=0.47]{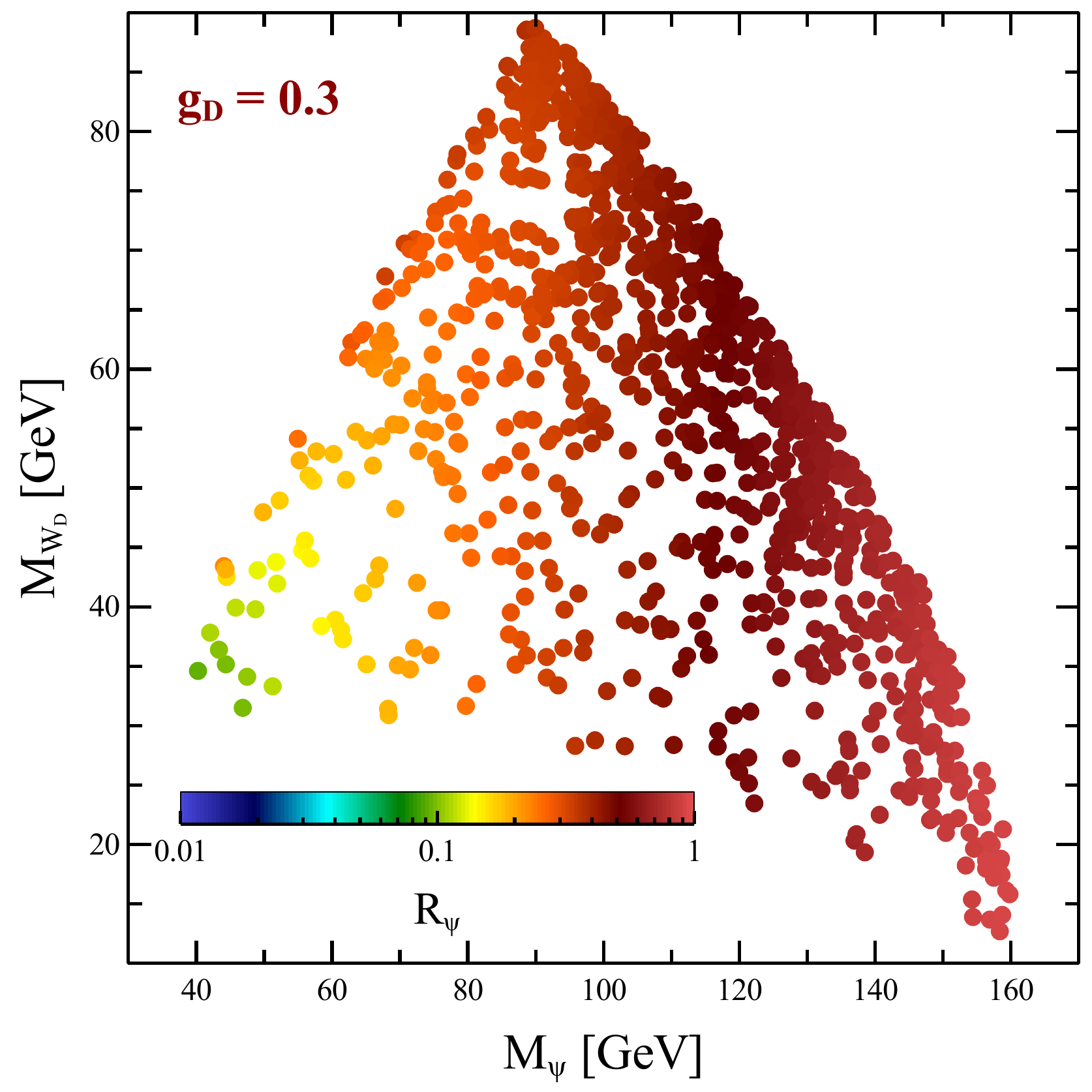}\,
\caption{Scan plot for showing the parameter space in $M_\psi$ vs $M_{W_D}$ plane allowed from relic density and direct detection experiments. In the left panel the color bar is for the variation on $M_{h_D}$ and in the right panel it is for relative contribution to the relic density coming from $\psi$ $\left(R_\psi = \Omega_{\psi}/\Omega_{\rm DM} \right)$.} 
\label{fig:relic:para}
\end{figure}

\begin{figure}
    \centering
\includegraphics[scale=0.47]{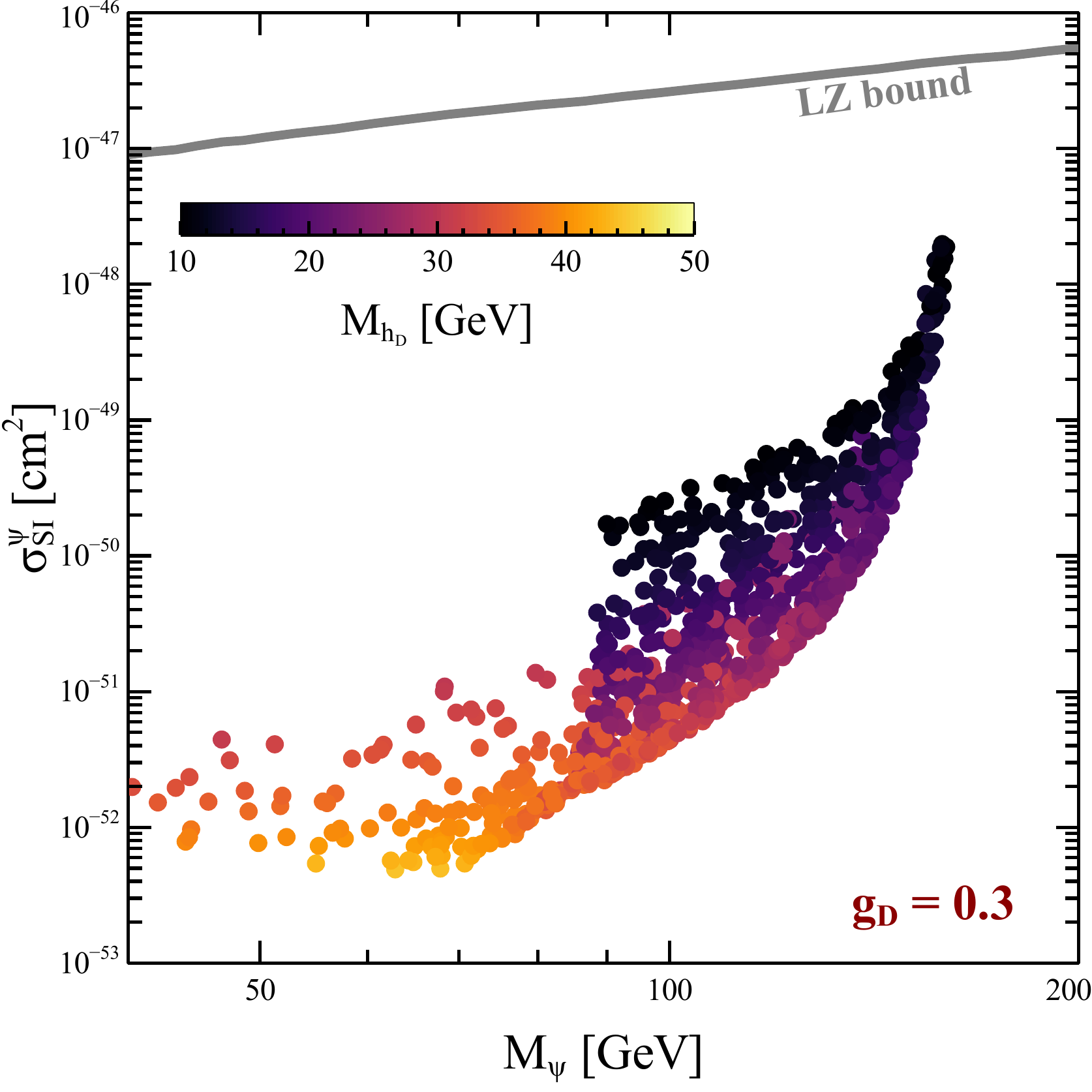}\,
\includegraphics[scale=0.47]{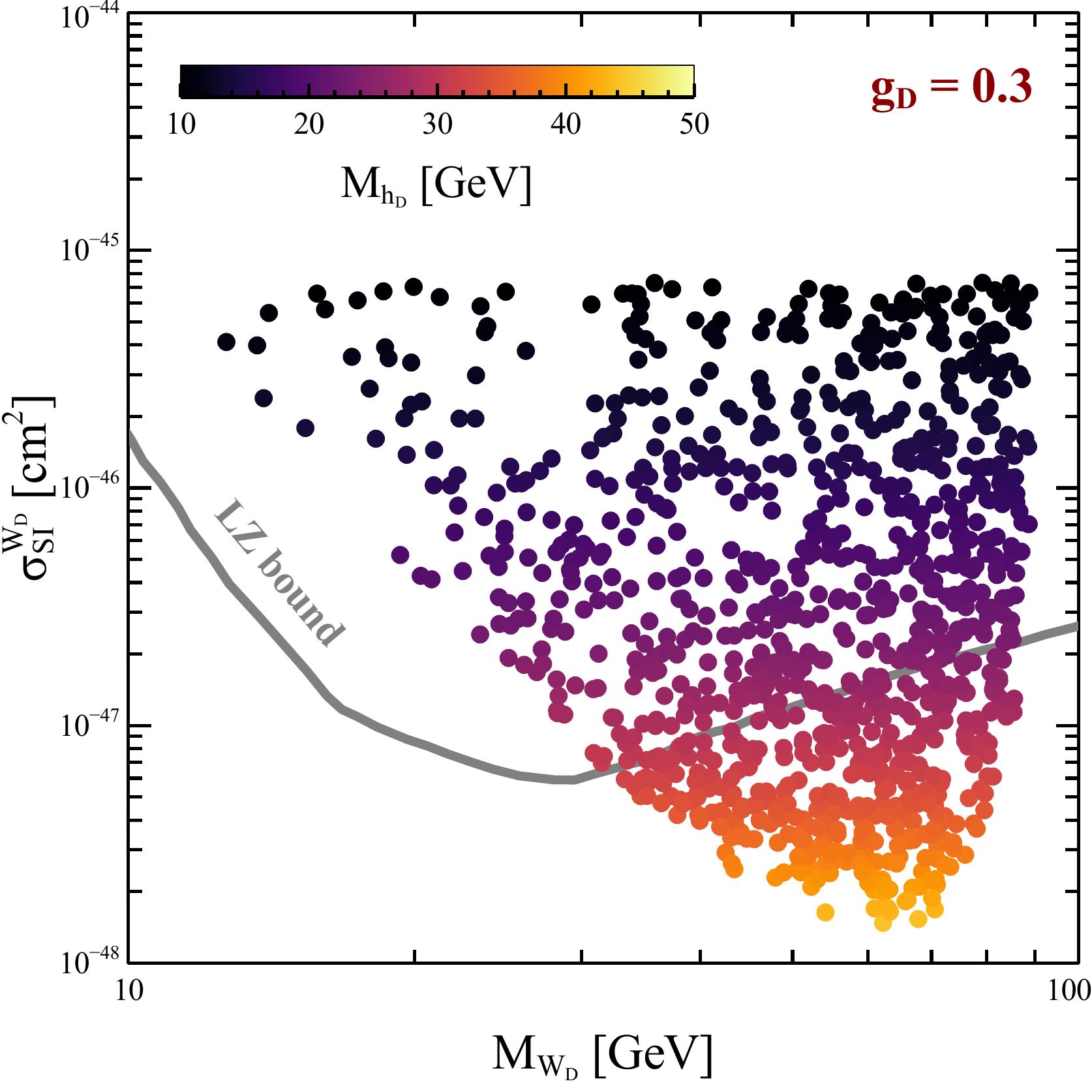}
    \caption{Left panel: Spin-independent direct detection cross-section as a function of $\psi$ mass. Right panel: Spin-independent direct detection cross-section as a function of $W_D$ mass.}
    \label{fig:DD:psi}
\end{figure}

\begin{figure}
    \centering
\includegraphics[scale=0.47]{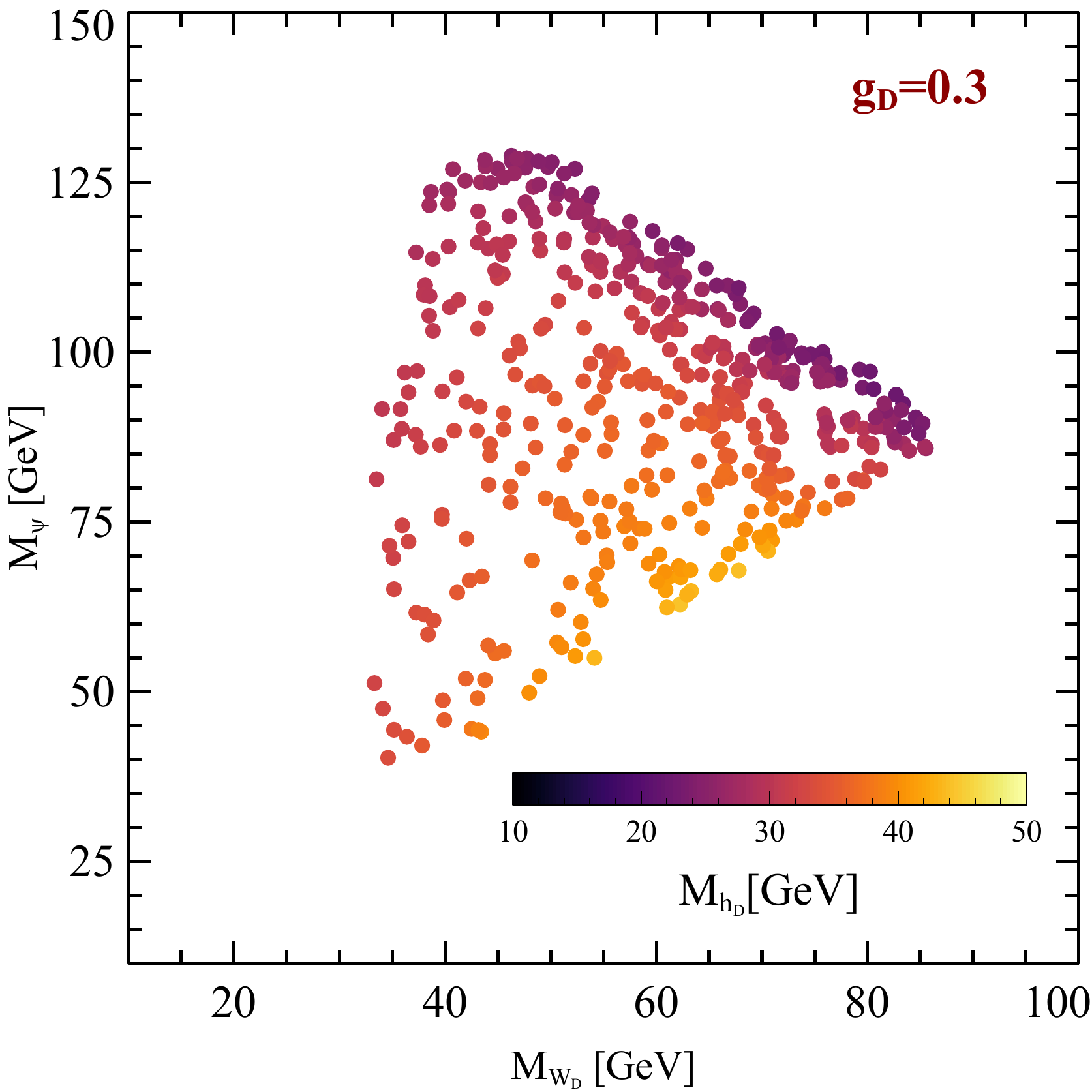}
    \caption{Final allowed parameter space in the plane of DM masses for $g_D=0.3$.}
    \label{fig:summary}
\end{figure}

The spin-independent direct detection cross-section for $\psi$-nucleon elastic scattering at one-loop level can be written as,
\begin{equation}
    \sigma_{\rm SI}^{\psi} = \frac{f_{h_D}^2 \mu_{\psi n}^2 m_n^2}{\pi M_{h_D}^4 v^2} f_p^2 \xi^2,
\end{equation}
where $f_{h_D}$ is the one-loop effective coupling of $\psi$ with dark Higgs with $\mu_{\psi n }$ being the reduced mass. In Fig.~\ref{fig:relic:para}, we have shown the allowed parameter space in $M_\psi - M_{W_D}$ plane from the observable total relic density constraints of dark matter. We have restricted our analysis in the mass range where $M_{h_D}<\,M_{W_D}<\,M_{\psi}$. We have also shown the effect of $M_{h_D}$ through color coding (left panel) and for this analysis we have kept $g_D$ to be fixed at 0.3. One can notice that, in this present scenario, the correct relic density can only be satisfied for a small region of parameter space. We have checked that for smaller values of dark gauge coupling ($g_D$) by one order of magnitude, thermal dark matter will always be overproduced in the present setup. The right panel of Fig.~\ref{fig:relic:para} shows the same parameter space with color code indicating the relative contribution of $\psi$ to total dark matter relic density. Clearly, for most part of the parameter space, $\psi$ dominates the total dark matter relic density as noticed in discussions above. Interestingly, some part of the parameter space we have shown above is already ruled out from the present limit of direct detection experiments, keeping the future detection prospects very promising. In Fig.~\ref{fig:DD:psi}, we have shown the spin-independent direct detection cross-section of for both the dark matter components as function of their masses. To calculate the effective spin-independent direct detection cross-section, we have multiplied the individual DM-nucleon scattering rate with the relative number densities of the dark matter particles. As can be seen by comparing with the latest upper limit on spin-independent dark matter nucleon cross-section \cite{LUX-ZEPLIN:2022qhg}, some part of the parameter space is excluded and the future experiments will be able to probe the parameter space. The scalar mixing is assumed to be $0.001$ in this case. While the one-loop direct detection cross-section for $\psi$ remains suppressed (as in the left panel of Fig.~\ref{fig:DD:psi}), the tree level cross-section of $W_D$ can be substantially large, in spite of its sub-dominant relic density. Since the direct search limit is on total DM relic density, some part of parameter space in $\psi$ mass plane will also be ruled out. The final allowed parameter space from relic density and direct detection constraints is shown in Fig.~\ref{fig:summary}. Since there are limited annihilation processes and $M_{W_D} < M_{\psi}$, the DM parameter space is limited to a small parameter space for fixed $g_D$. Since $\psi$ dominantly annihilates only into lighter DM $W_D$, its relic density is decided primarily by $g_D$ and $M_\psi$. The lighter DM relic density can be subsequently fixed by appropriately choosing dark scalar parameters. We also check the indirect detection bounds from gamma ray searches \cite{MAGIC:2016xys, HESS:2022ygk} and find the parameter space shown in Fig.~\ref{fig:summary} to be allowed for chosen value of scalar mixing.

Another interesting aspect of such light Dirac neutrino model is the enhancement of the effective relativistic degrees of freedom $N_{\rm eff}$ which can be probed at cosmic microwave background (CMB) experiments. The current $2\sigma$ limit on $N_{\rm eff}$ from the PLANCK experiment $N_{\rm eff}= 2.99^{+0.34}_{-0.33}$ \cite{Aghanim:2018eyx}, consistent with the SM prediction $N^{\rm SM}_{\rm eff}=3.045$. Future CMB experiment CMB Stage IV (CMB-S4) is expected reach a much better sensitivity of $\Delta {\rm N}_{\rm eff}={\rm N}_{\rm eff}-{\rm N}^{\rm SM}_{\rm eff}
= 0.06$ \cite{Abazajian:2019eic}, taking it closer to the SM prediction. Some recent works on light Dirac neutrinos and enhancement of $N_{\rm eff}$ can be found in \cite{Abazajian:2019oqj, FileviezPerez:2019cyn, Nanda:2019nqy, Han:2020oet, Luo:2020sho, Borah:2020boy, Luo:2020fdt, Mahanta:2021plx, Biswas:2021kio, Borah:2022obi}. In our model, the right chiral part of Dirac neutrino can get thermalised by virtue of its Yukawa interactions. Assuming all three $\nu_R$ to get thermalised in the early universe and decouple instantaneously above the electroweak scale, simple entropy conservation arguments lead to $\Delta N_{\rm eff} \approx 0.14$ \cite{Abazajian:2019oqj}, well within the reach of CMB-S4. Since our DM analysis does not depend on neutrino Yukawa couplings, they can be tuned appropriately to satisfy light neutrino mass criteria while guaranteeing $\nu_R$ thermalisation in the early universe. A detailed analysis of $N_{\rm eff}$ in this model is beyond the scope of this present work and can be found elsewhere in the context of similar radiative Dirac seesaw models.

\begin{figure}[htb]
 \vspace*{-5cm}
 \hspace*{-3cm}
 \includegraphics[scale=1.0]{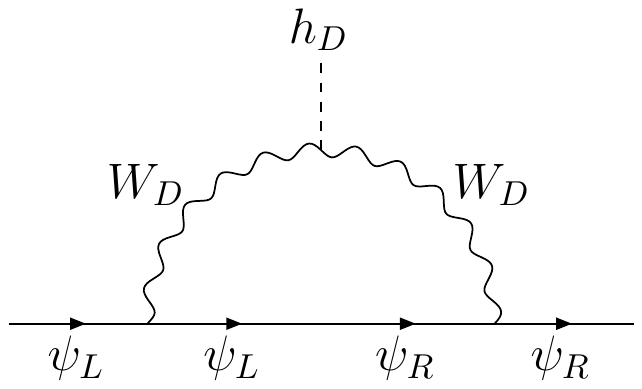}
 \vspace*{-21.5cm}
 \caption{Decay of $h_D$ to dark matter.}
 \label{fig:fimp}
 \end{figure}

\noindent \underline{\it Freeze-In Scenario}~:~
Under $SU(2)_D$, $\zeta$ cannot couple to $\psi_{L,R}$.  However, after 
the spontaneous breaking of $SU(2)_D$, the one dark Higgs boson $h_D$ 
obtains a connection to $\bar{\psi}_L \psi_R$ in one loop through the 
massive dark gauge bosons $W_D$, as shown in Fig. \ref{fig:fimp}.

In this scenario, $\psi$ is assumed to be light, of order GeV, and $W_D$ 
very heavy.  The effective $h_D \bar{\psi} \psi$ coupling is then
\begin{equation}
f_{h_D} = {3g_D^4 v_D M_\psi \over 8 \sqrt{2} \pi^2 M^2_{W_D}}.
\end{equation}
Assuming that $h_D$ is much heavier than $h$, then $h$ decays to 
$\psi \bar{\psi}$ through $h-h_D$ mixing.  Hence 
$f_h = f_{h_D}(2\lambda_3 v v_D/M^2_{h_D})$. 
The decay rate of $h \to \psi \bar{\psi}$ is
\begin{equation}
\Gamma_{h} = {f_{h}^2 m_{h} \over 8 \pi} \sqrt{1-4z^2}(1-2z^2),
\end{equation}
where $z=M_\psi/m_{h}$.  The correct dark matter relic density is 
obtained~\cite{Arcadi:2013aba} if $f_{h} \sim 10^{-12}z^{-1/2}$.  Assuming 
$M_\psi \sim 2$ GeV, $\lambda_3 \sim 10^{-4}$, and using 
$M_{W_D} = g_D v_D/\sqrt{2}$, this is satisfied for $M_{h_D}/g_D \sim 20$ TeV.
If the reheat temperature of the Universe is a few TeV, then the only 
production mechanism of $\psi$ is through $h$ decay.  Its relic 
denisty builds up to its present value until $h$ goes out of thermal equilibrium with the other SM particles.  It is known as a feebly interacting 
massive particle (FIMP)~\cite{Hall:2009bx}.

\begin{figure}[htb!]
\includegraphics[scale=0.7]{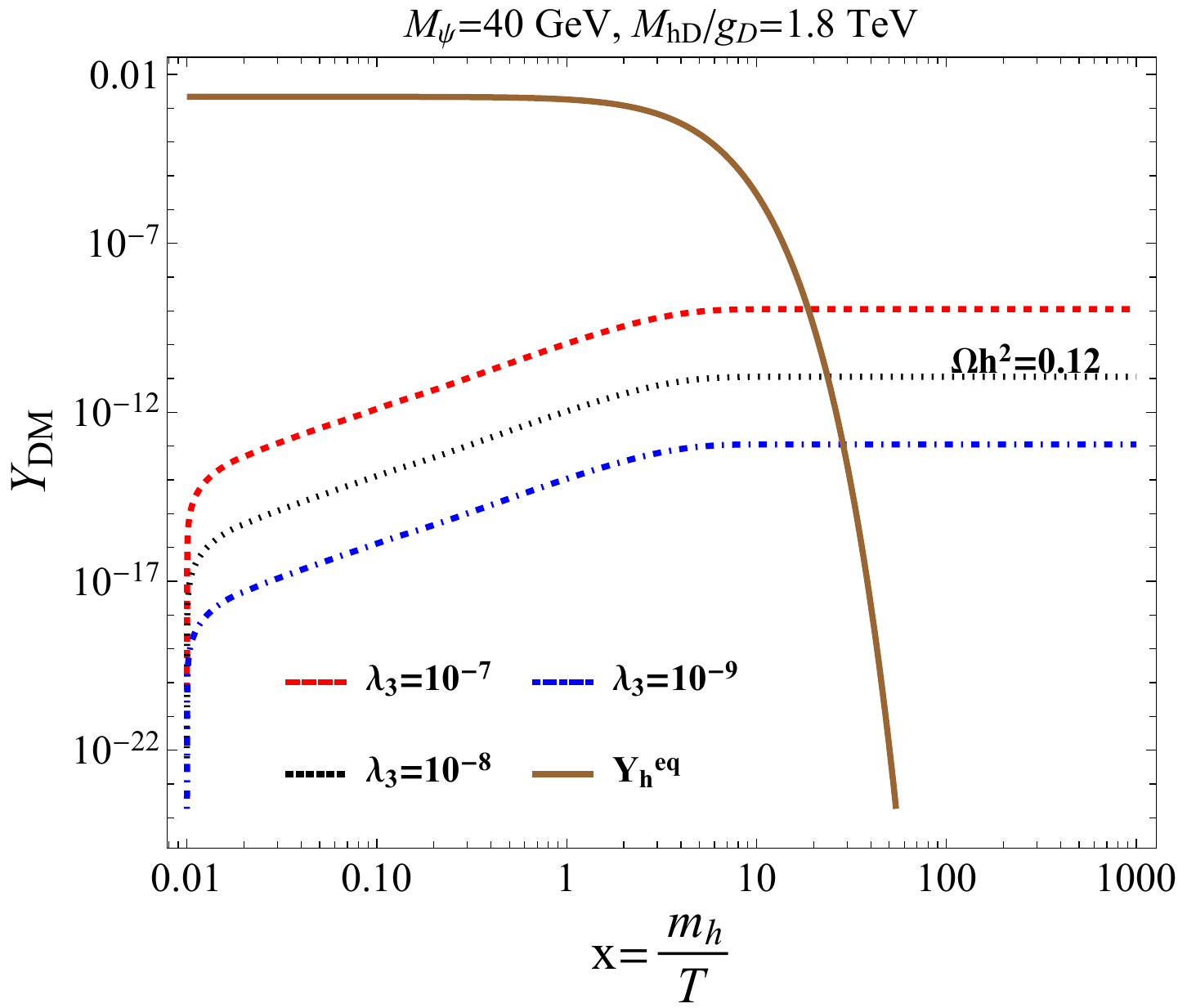}
\caption{Evolution of the comoving number density of dark matter as function of temperature.} 
\label{fig:relic:fimp}
\end{figure}

The Boltzmann equation for FIMP dark matter $\psi$, in terms of its comoving number density, can be written as
\begin{equation} \label{case1_psi_1}
    \frac{dY_{\psi}}{dx} = \frac{2 \beta}{x H } \Gamma_{h} \frac{K_1(x)}{K_2(x)} Y_{h}^{\rm eq},
\end{equation}
where $x=m_h/T$ and $K_i$ is modified Bessel function of i-th order. In Fig.~\ref{fig:relic:fimp}, we have shown the non-thermal production of dark matter comoving number density as a function of temperature for different benchmark values of the other parameters as shown in the figure. The benchmark values of $\lambda_3$ are chosen in a way that keeps the dark sector out of equilibrium with thermal bath. 

\noindent \underline{\it Concluding Remarks}~:~
The dark sector is postulated to consist of particles transforming under the 
gauge symmetry $SU(2)_D$, which is broken by a scalar doublet so that a 
global $SU(2)$ symmetry remains.  Adding fermions and scalars, also 
transforming as $SU(2)_D$ doublets, two Dirac neutrino masses may be 
obtained radiatively with dark matter in the loop.  The structure of the 
dark sector allows naturally two dark matter components in a thermal 
freeze-out scenario.  With a different choice of mass parameters, 
freeze-in production of dark matter may also be realized through Higgs 
decay.

\noindent \underline{\it Acknowledgement}~:~
This work was supported in part by the U.~S.~Department of Energy Grant 
No. DE-SC0008541.  


\end{document}